# A Novel Algorithm for Clustering of Data on the Unit Sphere via Mixture Models

Hien D. Nguyen*

**Abstract**

A new maximum approximate likelihood (ML) estimation algorithm for the mixture of Kent distribution is proposed. The new algorithm is constructed via the BSLM (block successive lower-bound maximization) framework and incorporates manifold optimization procedures within it. The BSLM algorithm is iterative and monotonically increases the approximate log-likelihood function in each step. Under mild regularity conditions, the BSLM algorithm is proved to be convergent and the approximate ML estimator is proved to be consistent. A Bayesian information criterion-like (BIC-like) model selection criterion is also derive, for the task of choosing the number of components in the mixture distribution. The approximate ML estimator and the BIC-like criterion are both demonstrated to be successful via simulation studies. A model-based clustering rule is proposed and also assessed favorably via simulations. Example applications of the developed methodology are provided via an image segmentation task and a neural imaging clustering problem.

## 1  Introduction

Let $\mathbb{R}^{d+1}$ denote the set of $(d+1)$-dimensional real vectors. For $d \in \mathbb{N}$, the natural numbers, we can embed the set

$$\mathbb{S}^d = \left\{ \boldsymbol{x} \in \mathbb{R}^{d+1} : \|\boldsymbol{x}\| = 1 \right\}$$

in $\mathbb{R}^{d+1}$. We refer to $\mathbb{S}^d$ as the $d$-sphere. In this article, we will be concerned only with the case where $d = 2$.

Many scientific phenomena can be represented as elements on a sphere. For example, in meteorology, the location of climatic events as being elements on the sphere, as can the location of earthquakes and seismic activities in geology. In medicine, vector cardiogram activity can described in terms of near-planar orbits in $\mathbb{R}^3$, and in astronomy, the distribution of celestial bodies can be seen as spherical in nature. See Mardia and Jupp (2000, Sec. 1.4) for further examples.

Let $\boldsymbol{X}$ be a random variable with support in $\mathbb{S}^2$. Over the years, there have been numerous suggestions regarding models for the generative process of $\boldsymbol{X}$. The most famous of these models include the von Mises-Fisher distribution (Fisher, 1953), the Bingham distribution (Bingham, 1974), and the Fisher-Bingham distribution

*Department of Mathematics and Statistics, La Trobe University, Bundoora 3086, Victoria Australia

(Mardia, 1975); see Mardia and Jupp (2000, Ch. 9) for a catalog of distributions on the sphere.

In this article, we are particularly interested in the special case of the Fisher-Bingham distribution in $\mathbb{S}^2$. This case is known as the Kent distribution and was introduced in Kent (1982). The Kent distribution can be described via the density function

$$f_K(\boldsymbol{x}; \boldsymbol{\psi}) = C^{-1}(\beta, \kappa) \exp\left[\kappa\left(\boldsymbol{x}^\top \boldsymbol{\xi}_1\right) + \beta\left(\boldsymbol{x}^\top \boldsymbol{\xi}_2\right)^2 - \beta\left(\boldsymbol{x}^\top \boldsymbol{\xi}_3\right)^2\right], \quad (1)$$

where

$$C(\beta, \kappa) = 2\pi \sum_{i=1}^{\infty} \frac{\Gamma(i+1/2)}{\Gamma(i+1)} \beta^{2i} \left(\frac{\kappa}{2}\right)^{-2i-1/2} I_{2i+1/2}(\kappa), \quad (2)$$

and $I_v(\kappa)$ is the modified Bessel function (Olver *et al.*, 2010, Ch. 10) with argument $v \in \mathbb{R}$, evaluated at $\kappa$. The parameter elements $\beta$ and $\kappa$ are restricted by the inequality $0 \leq 2\beta < \kappa$ and the matrix

$$\boldsymbol{\Xi} = \begin{bmatrix} \boldsymbol{\xi}_1 & \boldsymbol{\xi}_2 & \boldsymbol{\xi}_3 \end{bmatrix}$$

is restricted to be an orthonormal $3 \times 3$ real matrix. We put the parameter elements $\beta$, $\kappa$, and $\boldsymbol{\Xi}$ into the parameter vector $\boldsymbol{\psi} \in \Psi$, and we denote matrix transposition by $(\cdot)^\top$. Here, $\Psi$ is the parameter space under the stated restrictions.

As with modeling of data in $\mathbb{R}^{d+1}$, it is often interesting to consider random variables that arise from populations made up of heterogeneous subpopulations. Such population structures can be considered under the framework of finite mixture models (McLachlan and Peel, 2000) and lend themselves naturally to the conduct of clustering. The conduct of clustering using a generative finite mixture model is often referred to in the literature as model-based clustering.

The model-based clustering of data in $\mathbb{S}^2$ has been considered by numerous authors in the recent literature. For example, Banerjee *et al.* (2005) and Yang *et al.* (2016) considered the use of von Mises-Fisher distributions, Yamaji and Sato (2011) considered the Bingham distributions, Sra and Karp (2013) considered the Watson distributions, and Franke *et al.* (2016) considered the angular Gaussian distributions. However, we are particularly concerned with the work of Peel *et al.* (2001) who utilize a mixture of Kent distributions to model heterogeneity in rock mass fracture patterns.

We now suppose that $\boldsymbol{X}$ arises from a mixture of Kent distributions and let $Z \in [g] = \{1, \ldots, g\}$ be a random variable whereby $\mathbb{P}(Z = z) = \pi_z$ ($z \in [g]$) and $\sum_{z=1}^{g} \pi_z = 1$. Here, we say that $g$ is the number mixture components. Further, suppose that the density of $\boldsymbol{X}$, conditioned on $Z = z$, can be written as $f(\boldsymbol{x}|Z = z) = f_K(\boldsymbol{x}; \boldsymbol{\psi}_z)$, where $\boldsymbol{\psi}_z \in \Psi$. By the law of total probability, the marginal density of $\boldsymbol{X}$ can then be given as

$$f(\boldsymbol{x}; \boldsymbol{\theta}) := \sum_{z=1}^{g} \pi_z f_K(\boldsymbol{x}; \boldsymbol{\psi}_z). \quad (3)$$

We put all of the parameter elements (i.e. $\pi_z$ and $\boldsymbol{\psi}_z$, for each $z \in [g]$) into the vector $\boldsymbol{\theta} \in \Theta$. Here, $\Theta$ denotes the parameter space that incorporates all of the stated restrictions.

Let $\mathbf{X}_n = \{\boldsymbol{X}_i\}_{i=1}^n$ be an IID (independent and identically distributed) sample of $n$ observations from a distribution with density (3), and let $\mathbf{x}_n = \{\boldsymbol{x}_i\}_{i=1}^n$ be a realization of $\mathbf{X}_n$. Suppose that each $\boldsymbol{X}_i$ ($i \in [n]$) arises from a density $f(\boldsymbol{x}; \boldsymbol{\theta}_0)$, where $\boldsymbol{\theta}_0 \in \Theta$ is unknown. In the general mixture modeling context, given the realization $\mathbf{x}_n$, $\boldsymbol{\theta}_0$ can be estimated via maximum likelihood (ML) estimation. That is, to find an appropriate local maximizer to the log-likelihood function

$$L(\boldsymbol{\theta}; \mathbf{x}_n) = \sum_{i=1}^n \log f(\boldsymbol{x}_i; \boldsymbol{\theta}). \qquad (4)$$

Unfortunately, the functional form (1) involves the normalizing constant (2), which contains both an infinite series expansion and a modified Bessel function evaluation. This makes (4) difficult to maximize, beyond the usual log-sum-exp form of a mixture likelihood function.

In Peel *et al.* (2001), the authors instead utilize an approximate log-likelihood function of the form

$$\begin{aligned}\tilde{L}(\boldsymbol{\theta}; \mathbf{x}_n) &= \sum_{i=1}^n \log \tilde{f}(\boldsymbol{x}_i; \boldsymbol{\theta}) \\ &= \sum_{i=1}^n \log \sum_{z=1}^g \pi_z \tilde{f}_K(\boldsymbol{x}_i; \boldsymbol{\psi}_z) \end{aligned} \qquad (5)$$

where $\tilde{f}_K(\boldsymbol{x}; \boldsymbol{\psi})$ has form (1) with

$$\tilde{C}(\beta, \kappa) = \frac{2\pi \exp(\kappa)}{\sqrt{\kappa^2 - 4\beta^2}} \qquad (6)$$

in place of $C(\beta, \kappa)$. Here, the normalizing constant (6) is a large $\kappa$ asymptotic approximation of (2) and was given by Kent (1982). The authors sought to estimate the unknown parameter vector $\boldsymbol{\theta}_0$ by locating an appropriate maximizer of (5).

In their article, Peel *et al.* (2001) used an EM-like (expectation–maximization; Dempster *et al.*, 1977; see also McLachlan and Krishnan, 2008) algorithm for the iterative estimation of the unknown parameter vector. Their algorithm utilized a typical mixture model E-step in order to swap the "sum" and "log" in the log-sum-exp form of (5). Unfortunately, due to the constraints that each $\boldsymbol{\Xi}_z$ be in the space of orthonormal $3 \times 3$ matrices, for each $z \in [g]$. Due to this restriction, the authors utilized a weight method-of-moments estimator in place of the usual M-step of the EM algorithm, as moment expressions are known for Kent distributions (cf. Kent, 1982). Unfortunately, since the algorithm of Peel *et al.* (2001) is not a proper EM algorithm, it does not inherit the guarantees that EM algorithms provide. That is, does not provably increase the likelihood function at each iteration, nor does it guarantee convergence. Additionally since the "M-step" is not a proper maximization, it is not even guaranteed that the algorithm is searching for maximizer of (5), and thus may be producing an estimator that is unrelated to the objective altogether.

In this article, we seek to remedy the shortcomings of Peel *et al.* (2001). We do so by deriving a BSLM (block successive lower-bound maximization; Razaviyayn *et al.*, 2013) algorithm for the maximization of (5). Let $V_p(\mathbb{R}^q)$ denote the Stiefel manifold of orthonormal $p$-frames in $\mathbb{R}^q$ (cf. James, 1976). Upon noting that each $\Xi_z \in V_3\left(\mathbb{R}^3\right)$, we are able to overcome the previous difficulties preventing Peel *et al.* (2001) from conducting maximization in their M-step by utilizing modern techniques for optimization in Stiefel and other Riemannian manifolds; see Absil *et al.* (2008) for a recent introduction. In our BSLM algorithm, we utilize the recent ManifoldOptim (Martin *et al.*, 2016) for the R programming language (R Core Team, 2016), which is a wrapper for the ROPTLIB C++ library of Huang *et al.* (2016).

Since our algorithm is constructed under the BSLM framework, we are able to show that it monotonically increases the sequence of objective evaluates (i.e. evaluates of (5)) at each of its iterations. Furthermore, we are able to establish the convergence of accumulation points of the BSLM algorithm to its fixed points, under some mild regularity conditions.

By definition, the approximate ML estimator defined via the maximization of (5) is an extremum estimator (EE) in the sense of Amemiya (1985, Ch. 4). As such, we are able to prove a consistency theorem regarding its behavior for large $n$. Additionally, using some recent results of Baudry (2015), a variable selection theorem can also be proved under some mild regularity conditions that allows for the estimation of the number of components $g$ via a BIC-like (Bayesian information criterion; Schwarz, 1978) approach. To conclude the article, we present results from some simulation studies regarding the performance of our BSLM algorithm for computation of the approximate ML estimator. We also demonstrate how mixtures of Kent distributions can be used to cluster data on the unit sphere $\mathbb{S}^2$ and present applications of our methodology on some of real-world data sets.

The article proceeds as follows. The BSLM algorithm for approximate ML estimation of Kent distribution mixture models is derived in Section 2. Theoretical properties of the approximate ML estimator are presented in Section 3. Model-based clustering via Kent distributions mixtures is described and some simulation studies are present in Section 4. Demonstrations of real-world data applications are provided in Section 5. Conclusions are drawn in Section 6.

## 2 Approximate Maximum Likelihood Estimation

Let $\mathbf{X}_n$ be data arising from a mixture of Kent distributions of form (3), or a distribution that can be approximated by a distribution of form (3). Suppose that we observe a fixed realization $\mathbf{x}_n$ of $\mathbf{X}_n$, from which we can construct the approximate log-likelihood function (5).

We define $\hat{\boldsymbol{\theta}}_n$ to be the approximate ML estimator, or the extremum estimator (EE), in the terminology of Amemiya, 1985, Ch. 4), which we define as a suitable local-maximizer of the function (5). Due to the log-sum-exp form of (5) and due to the complexity of the manifold in which the parameter vector $\boldsymbol{\theta}$ resides, we cannot apply the usual first order condition (FOC) from elementary calculus in order to obtain the EE of (5). As such, we shall utilize a numerical scheme based upon the BSLM algorithm framework of Razaviyayn *et al.* (2013).

## 2.1 Block Successive Lower-Bound Maximization

Suppose that we wish to maximize some objective function $O(\boldsymbol{u})$, such that

$$\boldsymbol{u}^\top = \left(\boldsymbol{u}_1^\top, \ldots, \boldsymbol{u}_m^\top\right) \in \prod_{j=1}^m \mathbb{U}_j = \mathbb{U},$$

where $\mathbb{U}_j \subset \mathbb{R}^{p_j}$ for some $p_j \in \mathbb{N}$, for each $j \in [m]$ ($m \in \mathbb{N}$). Suppose also that $O(\boldsymbol{u})$ is difficult to optimize for some reason (e.g. the FOC are difficult to solve for, or the function is not differentiable).

For each $j \in [m]$, let $M_j(\boldsymbol{u}_j; \boldsymbol{v})$ be an easy to manipulate function with support $\mathbb{U}_j$, for fixed $\boldsymbol{v} \in \mathbb{U}$. We say that $M_j(\boldsymbol{u}_j; \boldsymbol{v})$ is a blockwise-minorizer of $O(\boldsymbol{u})$ in the $j$th block at $\boldsymbol{v}$ if it satisfies the assumptions: (A1) $M_j(\boldsymbol{v}_j; \boldsymbol{v}) = O(\boldsymbol{v})$ and (A2) $M_j(\boldsymbol{u}_j; \boldsymbol{v}) \leq O(\boldsymbol{w})$, where $\boldsymbol{w}^\top = \left(\boldsymbol{v}_1^\top, \ldots \boldsymbol{v}_{j-1}^\top, \boldsymbol{u}_j^\top, \boldsymbol{v}_{j+1}^\top, \ldots, \boldsymbol{v}_m^\top\right)$, for all $\boldsymbol{v}^\top = \left(\boldsymbol{v}_1^\top, \ldots, \boldsymbol{v}_m^\top\right) \in \mathbb{U}$.

To construct a BSLM algorithm, we firstly initialize it with some value $\boldsymbol{u}^{(0)}$. At the $r$th iteration of the algorithm ($r \in \mathbb{N}$), we then set $\boldsymbol{u}_j^{(r)}$ to be

$$\boldsymbol{u}_j^{(r)} = \begin{cases} \arg\max_{\boldsymbol{u}_j \in \mathbb{U}_j} M_j\left(\boldsymbol{u}_j; \boldsymbol{u}^{(r-1)}\right) & \text{if } j \in (r \bmod m) + 1, \\ \boldsymbol{u}_j^{(r-1)} & \text{otherwise,} \end{cases} \quad (7)$$

for each $j \in [m]$, where we then set $\boldsymbol{u}^{(r)\top} = \left(\boldsymbol{u}_1^{(r)\top}, \ldots, \boldsymbol{u}_m^{(r)\top}\right)$. Here, $r \bmod m$ yields the integer remainder of $r$ divided by $m$. We say that any algorithm that implements the iteration scheme (7) is a BSLM algorithm. Together, (A1), (A2), and scheme (7) can be combined to produce the sequence of inequalities

$$O\left(\boldsymbol{u}^{(r-1)}\right) \leq M_{(r \bmod m)+1}\left(\boldsymbol{u}_{(r \bmod m)+1}^{(r-1)}; \boldsymbol{u}^{(r-1)}\right) \quad (8)$$
$$\leq M_{(r \bmod m)+1}\left(\boldsymbol{u}_{(r \bmod m)+1}^{(r)}; \boldsymbol{u}^{(r-1)}\right)$$
$$\leq O\left(\boldsymbol{u}^{(r)}\right).$$

The sequence of inequalities (8) indicate that the sequence of iterates $\{\boldsymbol{u}^{(r)}\}$ that is produced by any BSLM algorithm will also generate a sequence of objective evaluates $\{O(\boldsymbol{u}^{(r)})\}$ that is monotonically increasing in $r$. This is a good result as it implies that any BSLM algorithm will only take steps towards a local maximum and will not decrease the objective of interest.

Define a generalized BSLM algorithm to be any algorithm for which

$$M_{(r \bmod m)+1}\left(\boldsymbol{u}_{(r \bmod m)+1}^{(r-1)}; \boldsymbol{u}^{(r-1)}\right) \leq M_{(r \bmod m)+1}\left(\boldsymbol{u}_{(r \bmod m)+1}^{(r)}; \boldsymbol{u}^{(r-1)}\right),$$

at each $r \in \mathbb{N}$. That is, a generalized algorithm need not solve strictly maximize the $[(r \bmod m) + 1]$th minorizer, at the $r$th iteration, but simply produce a solution that increases the minorizer. Such an algorithm does not satisfy the strict definition of a BSLM algorithm but will satisfy the inequalities of (8). This is a useful result as many numerical and iterative processes can be shown to be increasing but not necessarily globally optimal.

Consider an arbitrary algorithm, not necessarily a BSLM or a generalized BSLM algorithm, that is defined as follows. The algorithm starts at some initial value $\boldsymbol{u}^{(0)} \in \mathbb{U}$ (for some $\mathbb{U} \subset \mathbb{R}^p$; $p \in \mathbb{N}$) and it iterates via some scheme $\boldsymbol{u}^{(r)} \in \mathbb{M}\left(\boldsymbol{u}^{(r-1)}\right)$ for $r \in \mathbb{N}$, where $\mathbb{M}(\boldsymbol{u})$ is a point-to-set map from $\mathbb{U}$ into some subset of a nonempty subset of $\mathbb{U}$. Suppose further that there exists some continuous function $h(\boldsymbol{u})$, such that $h(\boldsymbol{v}) \geq h(\boldsymbol{u})$, where $\boldsymbol{v} \in \mathbb{M}(\boldsymbol{u})$.

Define a fixed point of the scheme $\mathbb{M}(\boldsymbol{u})$ to be some point $\{\boldsymbol{u}^{(\infty)}\} = \mathbb{M}\left(\boldsymbol{u}^{(\infty)}\right)$. We say that $\mathbb{M}(\boldsymbol{u})$ is strictly monotonic if $\boldsymbol{v} \in \mathbb{M}(\boldsymbol{u})$ implies $h(\boldsymbol{v}) > h(\boldsymbol{u})$ whenever $\boldsymbol{u} \neq \boldsymbol{u}^*$. Further say that the mapping $\mathbb{M}(\boldsymbol{u})$ is upper semicontinuous if the conditions $\boldsymbol{v}_k \in \mathbb{M}(\boldsymbol{u}_k)$ ($k \in \mathbb{N}$), and $\lim_{k \to \infty} \boldsymbol{u}_k \to \boldsymbol{u}$ and $\lim_{k \to \infty} \boldsymbol{v}_k \to \boldsymbol{v}$ imply that $\boldsymbol{v} \in \mathbb{M}(\boldsymbol{u})$. Lastly, we say that the mapping $\mathbb{M}(\boldsymbol{u})$ is uniformly compact on $\mathbb{U}$ if there exists a compact set $\mathbb{V}$ such that $\mathbb{M}(\boldsymbol{u}) \subset \mathbb{V}$ for all $\boldsymbol{u} \in \mathbb{U}$. It is notable that if $\mathbb{M}(\boldsymbol{u})$ returns a single value then continuity implies upper semicontinuity, and if $\mathbb{M}(\boldsymbol{u})$ is upper semicontinuous at $\boldsymbol{u}$ then it is also compact at $\boldsymbol{u}$. Denote the Euclidean norm of a vector $\boldsymbol{u}$ by $\|\boldsymbol{u}\|$. The following result regarding the global convergence of the sequence $\{\boldsymbol{u}^{(r)}\}$ is available from Meyer (1976); see also de Leeuw (1994).

**Theorem 1.** *Let $\mathbb{M}(\boldsymbol{u})$ be a point-to-set mapping such that $\mathbb{M}(\boldsymbol{u})$ is uniformly compact on $\mathbb{U}$, $\mathbb{M}(\boldsymbol{u})$ is upper semicontinuous on $\mathbb{U}$, and $\mathbb{M}(\boldsymbol{u})$ is strictly monotonic on $\mathbb{U}$ (with respect to the function $h(\boldsymbol{u})$). If $\{\boldsymbol{u}^{(r)}\}$ is a sequence that is generated by the iteration scheme $\boldsymbol{u}^{(r)} \in \mathbb{M}\left(\boldsymbol{u}^{(r-1)}\right)$, then all accumulation points of the algorithm will be fixed points, $h(\boldsymbol{u}) \to h(\boldsymbol{u}^*)$, where $\boldsymbol{u}^*$ is a fixed point, $\lim_{r \to \infty} \|\boldsymbol{u}^{(r)} - \boldsymbol{u}^{(r-1)}\| \to 0$, and either $\{\boldsymbol{u}^{(r)}\}$ converges to a limit point $\boldsymbol{u}^{(\infty)}$ or the accumulation points of $\{\boldsymbol{u}^{(r)}\}$ form a continuum.*

## 2.2 Algorithm Construction

For $\boldsymbol{u} = (u_1, \ldots, u_m)$ and $\boldsymbol{v} = (v_1, \ldots, v_m)$ with element $u_j > 0$ and $v_j > 0$ for each $j \in [m]$, Zhou and Lange (2010) proposed the minorizer

$$M(\boldsymbol{u}; \boldsymbol{v}) = \sum_{j=1}^{m} \frac{v_j}{\sum_{z=1}^{m} v_z} \log u_j - \sum_{j=1}^{m} \frac{v_j}{\sum_{z=1}^{m} v_z} \log \frac{v_j}{\sum_{z=1}^{m} v_z} \qquad (9)$$

in all $m$ coordinates, for $O(\boldsymbol{u}) = \log\left(\sum_{j=1}^{m} v_j\right)$. Applying (9) to (5) yields the minorizer at $\boldsymbol{\theta}^{(r-1)}$:

$$\begin{aligned} Q\left(\boldsymbol{\theta}; \boldsymbol{\theta}^{(r-1)}\right) &= \sum_{i=1}^{n} \sum_{z=1}^{g} \tau_z\left(\boldsymbol{x}_i; \boldsymbol{\theta}^{(r-1)}\right) \log \pi_z \qquad (10) \\ &+ \sum_{i=1}^{n} \sum_{z=1}^{g} \tau_z\left(\boldsymbol{x}_i; \boldsymbol{\theta}^{(r-1)}\right) \log \tilde{f}_K\left(\boldsymbol{x}_i; \boldsymbol{\psi}_z^{(r-1)}\right) \\ &+ \sum_{i=1}^{n} \sum_{z=1}^{g} \tau_z\left(\boldsymbol{x}_i; \boldsymbol{\theta}^{(r-1)}\right) \log \tau_z\left(\boldsymbol{x}_i; \boldsymbol{\theta}^{(r-1)}\right), \end{aligned}$$

where

$$\tau_z(\boldsymbol{x}; \boldsymbol{\theta}) = \pi_z \tilde{f}_K(\boldsymbol{x}; \boldsymbol{\psi}_z) / \sum_{j=1}^{g} \pi_j \tilde{f}_K(\boldsymbol{x}; \boldsymbol{\psi}_j)$$

for each $z \in [g]$.

We can expand (10) in order to obtain

$$Q\left(\boldsymbol{\theta}; \boldsymbol{\theta}^{(r-1)}\right) = \sum_{i=1}^{n} \sum_{z=1}^{g} \tau_z\left(\boldsymbol{x}_i; \boldsymbol{\theta}^{(r-1)}\right) \log \pi_z \qquad (11)$$

$$+ \frac{1}{2} \sum_{i=1}^{n} \sum_{z=1}^{g} \tau_z\left(\boldsymbol{x}_i; \boldsymbol{\theta}^{(r-1)}\right) \log\left(\kappa_z^2 - 4\beta_z^2\right)$$

$$+ \sum_{i=1}^{n} \sum_{z=1}^{g} \tau_z\left(\boldsymbol{x}_i; \boldsymbol{\theta}^{(r-1)}\right) \kappa_z \left[\left(\boldsymbol{x}_i^\top \boldsymbol{\xi}_1\right) - 1\right]$$

$$+ \sum_{i=1}^{n} \sum_{z=1}^{g} \tau_z\left(\boldsymbol{x}_i; \boldsymbol{\theta}^{(r-1)}\right) \beta_z \left[\left(\boldsymbol{x}_i^\top \boldsymbol{\xi}_2\right)^2 - \left(\boldsymbol{x}_i^\top \boldsymbol{\xi}_3\right)^2\right]$$

$$+ C\left(\mathbf{x}_n; \boldsymbol{\theta}^{(r-1)}\right),$$

where $C\left(\mathbf{x}_n; \boldsymbol{\theta}^{(r-1)}\right)$ is a constant that does not depend on the active parameter vector $\boldsymbol{\theta}$. We partition $\boldsymbol{\theta}$ into two coordinates $\boldsymbol{\theta}_1^\top = (\pi_1, \beta_1, \kappa_1, \ldots, \pi_g, \beta_g, \kappa_g)$ and $\boldsymbol{\theta}_2^\top = \left(\mathrm{vec}^\top \boldsymbol{\Xi}_1, \ldots, \mathrm{vec}^\top \boldsymbol{\Xi}_g\right)$, such that $\boldsymbol{\theta}^\top = (\boldsymbol{\theta}_1^\top, \boldsymbol{\theta}_2^\top)$. Here, $\mathrm{vec}(\cdot)$ extracts the unique elements of its matrix input. Using (11), we can minorize (5) with respect to the coordinate $\boldsymbol{\theta}_1$ via the blockwise-minorizer

$$Q_1\left(\boldsymbol{\theta}_1; \boldsymbol{\theta}^{(r-1)}\right) = Q\left(\begin{bmatrix} \boldsymbol{\theta}_1 & \boldsymbol{\theta}_2^{(r-1)} \end{bmatrix}; \boldsymbol{\theta}^{(r-1)}\right). \qquad (12)$$

We note that the problem of maximizing (12) with respect to $\pi_z$ is separable to that of maximizing the function $\beta_z$ and $\kappa_z$, for all $z \in [g]$. As such, we can solve these two problems separately and simultaneous. The maximization of (12) with respect to $\pi_z$, for each $z$, is a standard problem in mixture modeling (cf. McLachlan and Peel, 2000, Sec. 2.8) and has the well-known solution

$$\pi_z^* = n^{-1} \sum_{i=1}^{n} \tau_z\left(\boldsymbol{x}_i; \boldsymbol{\theta}^{(r-1)}\right), \qquad (13)$$

for each $z \in [g]$.

Notice that (12) is linearly separable with respect to the parameters belonging to each of the $g$ mixture components. Thus, we can consider the optimization over each of $\beta_z$ and $\kappa_z$ separately, for each $z \in [g]$. Now, make the following definitions: let

$$a_z^{(r-1)} = \frac{1}{2} \sum_{i=1}^{n} \tau_z\left(\boldsymbol{x}_i; \boldsymbol{\theta}^{(r-1)}\right),$$

$$b_z^{(r-1)} = \sum_{i=1}^{n} \tau_z\left(\boldsymbol{x}_i; \boldsymbol{\theta}^{(r-1)}\right) \left[\left(\boldsymbol{x}_i^\top \boldsymbol{\xi}_1^{(r-1)}\right) - 1\right],$$

and

$$c_z^{(r-1)} = \sum_{i=1}^{n} \tau_z\left(\boldsymbol{x}_i; \boldsymbol{\theta}^{(r-1)}\right) \left[\left(\boldsymbol{x}_i^\top \boldsymbol{\xi}_2^{(r-1)}\right)^2 - \left(\boldsymbol{x}_i^\top \boldsymbol{\xi}_3^{(r-1)}\right)^2\right],$$

for each $z \in [g]$.

We can maximize (11) with respect to $\beta_z$ and $\kappa_z$ for all $z \in [g]$ by solving each of the concave subproblems:

$$(\beta_z^*, \kappa_z^*) = \arg\max_{\beta_z, \kappa_z} \ a_z^{(r-1)} \log\left(\kappa_z^2 - 4\beta_z^2\right) + b_z^{(r-1)} \kappa_z + c_z^{(r-1)} \beta_z, \qquad (14)$$

under the restriction $0 \leq 2\beta_z < \kappa_z$, for each $z$. Let $\bar{B} > 0$ and $\bar{K} > 0$ be two arbitrarily small constants that are bounded away from zero. Assume further that, for each $z$, $\beta_z \geq \bar{B}$ and $\kappa_z - 2\beta_z \geq \bar{K}$. Under such restrictions, we can assure that the solutions to problems of form (14) are non-degenerate.

There are numerous available solvers that can be leveraged for such problems. We can therefore obtain efficient solutions to each of the subproblems (14). Thus, we have obtained an efficient numerical method for computing

$$\boldsymbol{\theta}_1^{(r)} = \arg\max_{\boldsymbol{\theta}_1} \ Q_1\left(\boldsymbol{\theta}_1; \boldsymbol{\theta}^{(r-1)}\right),$$

via the solution (13) and an efficient solver to problem (14). One such efficient solution framework is to utilize the sequential quadratic programming (SQP) approach of Nocedal and Wright (2006, Ch. 18).

Consider the minorizer

$$Q_2\left(\boldsymbol{\theta}_2; \boldsymbol{\theta}^{(r-1)}\right) = Q\left(\begin{bmatrix} \boldsymbol{\theta}_1^{(r-1)} & \boldsymbol{\theta}_2 \end{bmatrix}; \boldsymbol{\theta}^{(r-1)}\right), \qquad (15)$$

of (5) with respect to $\boldsymbol{\theta}_2$. Notice that (15) is linearly separable in the $g$ components and is not dependent on the first two rows of the expression. Thus, the problem of maximizing (15) with respect to each $\boldsymbol{\Xi}_z$, for $z \in [g]$, by obtaining the solution to the problems:

$$\boldsymbol{\Xi}_z^* = \arg\max_{\boldsymbol{\Xi}_z \in V_3(\mathbb{R}^3)} \sum_{i=1}^n \tau_z\left(\boldsymbol{x}_i; \boldsymbol{\theta}^{(r-1)}\right) \kappa_z^{(r-1)} \left[\left(\boldsymbol{x}_i^\top \boldsymbol{\xi}_1\right) - 1\right] \qquad (16)$$
$$+ \sum_{i=1}^n \tau_z\left(\boldsymbol{x}_i; \boldsymbol{\theta}^{(r-1)}\right) \beta_z^{(r-1)} \left[\left(\boldsymbol{x}_i^\top \boldsymbol{\xi}_2\right)^2 - \left(\boldsymbol{x}_i^\top \boldsymbol{\xi}_3\right)^2\right].$$

A solutions to each of the problems of form (16) may be found using the many manifold optimization solvers from the package Martin *et al.* (2016). For example, the implemented methods of Absil *et al.* (2007) and Huang *et al.* (2015) have found successful implementation on similar problems in the literature.

Assuming that a solution to (16) exists, for each $z \in [g]$, we can combine it, the solution (13), and an efficient solver to the problem problems (14) in order to obtain a BSLM algorithm via definition (7). Unfortunately, the existence of a solution to (16) cannot be guaranteed, and checking the solution is made difficult due to the complexity of geodesics of the Stiefel manifolds (cf. Edelman *et al.*, 1998). Thus, in such cases, we need only verify that the solver for the problem (16) obtains a solution $\boldsymbol{\theta}_2^{(r)}$ whereby

$$Q_2\left(\boldsymbol{\theta}_2^{(r)}; \boldsymbol{\theta}^{(r-1)}\right) > Q_2\left(\boldsymbol{\theta}_2^{(r-1)}; \boldsymbol{\theta}^{(r-1)}\right).$$

In such a case, we can utilize Theorem 1 in order to obtain the following convergence result.

**Proposition 1.** *Assume that the solvers for problems (14) and (16), along with solution (13), yield solutions $\boldsymbol{\theta}_1^{(r)}$ and $\boldsymbol{\theta}_2^{(r)}$ that fulfill conditions*

$$Q_1\left(\boldsymbol{\theta}_1^{(r)}; \boldsymbol{\theta}^{(r-1)}\right) > Q_1\left(\boldsymbol{\theta}_1^{(r-1)}; \boldsymbol{\theta}^{(r-1)}\right),$$

*and*

$$Q_2\left(\boldsymbol{\theta}_2^{(r)}; \boldsymbol{\theta}^{(r-1)}\right) > Q_2\left(\boldsymbol{\theta}_2^{(r-1)}; \boldsymbol{\theta}^{(r-1)}\right),$$

*respectively, at each iteration $r \in \mathbb{N}$. If $\{\boldsymbol{\theta}^{(r)}\}$ is the sequence of iterates obtained via the BSLM algorithm constructed from definition (7), solution (13), and the solvers to problems (14) and (16), then $\lim_{r\to\infty} \|\boldsymbol{\theta}^{(r)} - \boldsymbol{\theta}^{(r-1)}\| \to 0$, and either $\{\boldsymbol{\theta}^{(r)}\}$ converges to a limit point $\boldsymbol{\theta}^{(\infty)}$ or the accumulation points of $\{\boldsymbol{\theta}^{(r)}\}$ form a continuum.*

## 3 Properties of the Approximate Maximum Likelihood Estimator

Let $\Theta$ be the set of all valid parameter vector values, under the additional restriction that $\beta_z \geq \bar{B}$ and $\kappa_z - 2\beta_z \geq \bar{K}$, for each $z \in [g]$. We can check that $\mathbb{E}\sup_{\boldsymbol{\theta}\in\mathbb{B}} \log \tilde{f}(\boldsymbol{X}_i; \boldsymbol{\theta}) < \infty$ and that $\sup_{\boldsymbol{\theta}\in\mathbb{B}} \log \tilde{f}(\boldsymbol{x}_i; \boldsymbol{\theta})$ is measurable for every sufficiently small ball $\mathbb{B} \subset \Theta$. Consider the set

$$\Theta_0 = \left\{\boldsymbol{\theta}_0 \in \Theta: \; \mathbb{E}\log \tilde{f}(\boldsymbol{X}_i; \boldsymbol{\theta}_0) = \sup_{\boldsymbol{\theta}\in\Theta} \mathbb{E}\log \tilde{f}(\boldsymbol{X}_i; \boldsymbol{\theta})\right\}.$$

Using van der Vaart (1998, Thm. 5.14), we obtain the following result regarding the EE estimator $\hat{\boldsymbol{\theta}}_n$.

**Proposition 2.** *If $\hat{\boldsymbol{\theta}}_n \in \Theta$ satisfies the condition $\tilde{L}\left(\hat{\boldsymbol{\theta}}_n; \mathbf{X}_n\right) \geq \tilde{L}\left(\boldsymbol{\theta}_0; \mathbf{X}_n\right) - o_{\mathbb{P}}(n)$ (for each $n \in \mathbb{N}$) for some $\boldsymbol{\theta}_0 \in \Theta_0$, then for every $\epsilon > 0$ and compact subset $\mathbb{K} \subset \Theta$, we have*

$$\lim_{n\to\infty} \mathbb{P}\left(\inf_{\boldsymbol{\theta}\in\Theta_0} \|\hat{\boldsymbol{\theta}}_n - \boldsymbol{\theta}\| \geq \epsilon \text{ and } \hat{\boldsymbol{\theta}}_n \in \mathbb{K}\right) = 0.$$

Proposition 2 is a useful result as it allows for the existence of multiple global maximizers of the approximate log-likelihood function (5). In the language of Amemiya (1985), Proposition 2 states that there exists a consistent maximizer of (5). The proposition does not provide a guide regarding how one should choose among the maximizers. Two suggestions from Amemiya (1985) are to consider whether or not the maximizer of choice makes sense from a scientific perspective, and whether or not the maximizer is stable when estimation is performed via numerical means with different initializations. The problem of initialization of algorithms for mixture-type models has been broadly studied in the literature; see for example McLachlan (1988).

Thus far, we have considered the number of mixture components $g$ to be fixed. However, in reality, it is an unknown that also requires estimation from data. In mixture model problems, it is common to utilize information criteria (IC) such as the Akaike information criterion (AIC; Akaike, 1974) or the BIC for selecting among different values of $g$. The following theorem of Baudry (2015) is useful for constructing IC and validating theoretically validating the large sample performance of the constructed IC.

**Theorem 2.** *Let $O(\boldsymbol{\theta}; \mathbf{X}_n) = \sum_{i=1}^{n} o(\boldsymbol{\theta}; \boldsymbol{X}_i)$ be a random objective function that is composed of $n \in \mathbb{N}$ objectives of the form $o(\boldsymbol{\theta}; \boldsymbol{X}_i)$, for $i \in [n]$. Define $\{\Theta_k\}_{k=1}^{G}$ to be a collection of models, such that $\Theta_k \subset \mathbb{R}^{d_k}$ for each $k \in [G]$, and for $G \in \mathbb{N}$ and $d_1 \leq \cdots \leq d_G$. Further define*

$$\Theta_k^0 = \left\{ \boldsymbol{\theta}^{[k]} : \mathbb{E} \, o\left(\boldsymbol{\theta}^{[k]}; \boldsymbol{X}\right) = \max_{\boldsymbol{\theta} \in \Theta_k} \mathbb{E} \, o(\boldsymbol{\theta}; \boldsymbol{X}) \right\}.$$

*Make the following assumptions: (A1)*

$$g_0 = \min\left[\arg\max_{k \in [G]} \mathbb{E} \, o\left(\boldsymbol{\theta}_0^{[k]}; \boldsymbol{X}\right)\right],$$

*where $\boldsymbol{\theta}_0^{[k]} \in \Theta_k^0$; (A2) $\hat{\boldsymbol{\theta}}_n^{[k]} \in \Theta_k$ is such that $O(\boldsymbol{\theta}; \mathbf{X}_n) \geq O(\boldsymbol{\theta}; \mathbf{X}_n) + o_\mathbb{P}(n)$ and $n^{-1} O(\boldsymbol{\theta}; \mathbf{X}_n)$ converges in probability to $\mathbb{E} \, o\left(\boldsymbol{\theta}_0^{[k]}; X\right)$, for all $k \in [G]$; (A3) the function $\text{pen}_n(k)$ is such that $\text{pen}_n(k) > 0$ and $\text{pen}_n(k) = o_\mathbb{P}(n)$ as $n \to \infty$, and $\text{pen}_n(k) - \text{pen}_n(l)$ diverges in probability, whenever $l < k$; and (A4)*

$$O\left(\hat{\boldsymbol{\theta}}_n^{[k]}; \mathbf{X}_n\right) - O\left(\hat{\boldsymbol{\theta}}_n^{[g_0]}; \mathbf{X}_n\right) = O_\mathbb{P}(1)$$

*for any*

$$k \in \left\{ l \in [G] : \mathbb{E} o\left(\boldsymbol{\theta}_0^{[l]}; \boldsymbol{X}\right) = \max_{l \in [G]} \mathbb{E} o\left(\boldsymbol{\theta}_0^{[l]}; \boldsymbol{X}\right) \right\}.$$

*If Assumptions (A1)–(A4) are fulfilled and*

$$\hat{g}_n = \arg\min_{k \in [G]} \left[ -O\left(\hat{\boldsymbol{\theta}}_n^{[k]}; \mathbf{X}_n\right) + \text{pen}_n(k) \right],$$

*then $\lim_{n \to \infty} \mathbb{P}(\hat{g}_n \neq g_0) = 0$.*

Consider the BIC penalty of the form $\text{pen}_n(k) = \log n \times \dim\left(\hat{\boldsymbol{\theta}}_n^{[k]}\right)/2$, where $\dim(\cdot)$ returns the dimension of the vector input. We can check that the BIC penalty fulfills the conditions of (A3) from Theorem 2 by noting that $\log n > 0$ diverges to infinity and that $\lim_{n \to \infty} n^{-1} \log n = 0$. Next, Baudry (2015, Lem. 8.1) states that (A2) can be fulfilled, for fixed $k \in [G]$ provided that $n^{-1} O(\boldsymbol{\theta}; \mathbf{X}_n)$ converges in probability uniformly to $\mathbb{E} \, o(\boldsymbol{\theta}; X)$. For compact subsets of Euclidean spaces and IID data, we obtain such results via uniform laws of large numbers such as the classic result of Jennrich (1969). Assumption (A1) simply declares that we are searching for the most parsimonious model among equally well-fitting alternatives and (A4) must be taken as given unless more stringent assumptions are placed on the objective $O(\boldsymbol{\theta}; \mathbf{X}_n)$ (cf. Baudry, 2015, Appendix 8.3). Using Theorem 2, we obtain the following result regarding a BIC-like criterion for selecting the number of mixture components $g$ in the context of approximate ML estimation of Kent distribution mixtures.

**Proposition 3.** *Let $\mathbf{X}_i$ be an IID random sample. Define $\{\Theta_k\}_{k=1}^{G}$ to be a collection of parameter spaces defined by $k \in [G]$ component Kent distribution mixtures of form (3) with approximation (6), for $G \in \mathbb{N}$. Further define*

$$\Theta_k^0 = \left\{ \boldsymbol{\theta}^{[k]} : \mathbb{E} \log \tilde{f}\left(\boldsymbol{X}; \boldsymbol{\theta}^{[k]}\right) = \max_{\boldsymbol{\theta} \in \Theta_k} \mathbb{E} \log \tilde{f}(\boldsymbol{X}; \boldsymbol{\theta}) \right\}.$$

*Make the following assumptions: (B1)*

$$g_0 = \min \left[ \arg \max_{k \in [G]} \mathbb{E} \log \tilde{f} \left( \boldsymbol{X}; \boldsymbol{\theta}_0^{[k]} \right) \right],$$

*where $\boldsymbol{\theta}_0^{[k]} \in \Theta_k^0$ and (B2)*

$$\tilde{L} \left( \hat{\boldsymbol{\theta}}_n^{[k]}; \mathbf{X}_n \right) - \tilde{L} \left( \hat{\boldsymbol{\theta}}_n^{[g_0]}; \mathbf{X}_n \right) = O_\mathbb{P}(1)$$

*for any*

$$k \in \left\{ l \in [G] : \mathbb{E} \log \tilde{f} \left( \boldsymbol{X}; \boldsymbol{\theta}_0^{[l]} \right) = \max_{l \in [G]} \mathbb{E} \log \tilde{f} \left( \boldsymbol{X}; \boldsymbol{\theta}_0^{[l]} \right) \right\}.$$

*If Assumptions (B1) and (B2) are fulfilled, and*

$$\hat{g}_n = \arg \min_{k \in [G]} \left[ -\tilde{L} \left( \hat{\boldsymbol{\theta}}_n^{[k]}; \mathbf{X}_n \right) + pen_n(k) \right], \tag{17}$$

*where $pen_n(k) = (11g/2) \log n$ then $\lim_{n \to \infty} \mathbb{P}(\hat{g}_n \neq g_0) = 0$.*

We note that in Proposition 3, the penalty is obtained via the BIC penalty formula (i.e. $\dim \left( \hat{\boldsymbol{\theta}}_n^{[k]} \right) = 11g$). The proposition allows us to use a BIC-like IC to select the number of mixture components $\hat{g}_n$ that provides the most parsimonious fit to the data. This result extends upon the well known IC results for mixture models of Leroux (1992) and Keribin (2000).

## 4 Simulation Studies

We perform a set of simulation studies in order to assess the performance of the BSLM algorithm that is presented in Section 2 and to check the theoretical claims that are presented in Section 3. All of our computations are performed within the R programming environment (R Core Team, 2016) on a MacBook Pro with a 2.2 GHz Intel Core i7 processor, 16 GB of 1600 MHz DDR3 memory, and a 500 GB SSD. We utilize the the SQP algorithm of Chen and Yin (2017) to solve the subproblems of form (14). The manifold optimization algorithms of Martin *et al.* (2016) are used to solve the subproblems of form (16). Each time the BSLM algorithm is applied for the computation of the EE $\hat{\boldsymbol{\theta}}_n$, we set the bounds to be $\bar{B} = \bar{K} = 10^{-5}$ and run the algorithm for 100 iterations.

### 4.1 Small-Sample Accuracy of the Approximate Maximum Likelihood Estimator

Proposition 2 provides a theoretical guarantee that the sequence of EEs obtained via maximization of (5) will converge to a parameter vector that is related generating process of the data via the maximizer of the expectation of (5), as the size of the data set $n$ gets large. However, the proposition does not guarantee that the parameter vector that maximizes the average of the approximation (5) is the same as that which underlies the data generating process, nor does it guarantee that the EEs will converge towards the generative parameter values for some finite sample size $n$. Our first pair of simulation studies (S-I) and (S-II) are designed to assess the accuracy of the EEs with respect to these two outstanding questions.

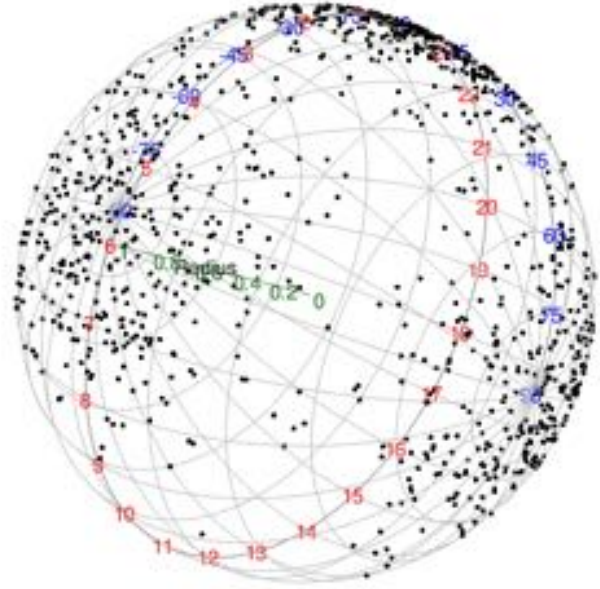

Figure 1: A single instance of an $n = 1000$ observations sample that is simulated under the $g = 3$ component mixture scenario of S-I.

In both S-I and S-II, we generate $n = 1000$ observations from mixtures of von Mises-Fisher distributions; that is, mixtures having densities of the form (1) with $\beta = 0$. In S-I we simulate the data from a $g = 3$, where $\kappa_z = 10$ and $\pi_z = 1/3$ for all $z \in [g]$, and $\boldsymbol{\xi}_{11}^\top = (1,0,0)$, $\boldsymbol{\xi}_{21}^\top = (0,1,0)$, and $\boldsymbol{\xi}_{31}^\top = (0,0,1)$, with equal probability. In S-II we simulate the data from a $g = 6$, where $\kappa_z = 20$ and $\pi_z = 1/6$ for all $z \in [g]$, and $\boldsymbol{\xi}_{11}^\top = (-1,0,0)$, $\boldsymbol{\xi}_{21}^\top = (0,-1,0)$ $\boldsymbol{\xi}_{31}^\top = (0,0,-1)$, $\boldsymbol{\xi}_{41}^\top = (1,0,0)$, $\boldsymbol{\xi}_{51}^\top = (0,1,0)$, and $\boldsymbol{\xi}_{61}^\top = (0,0,1)$. Note that we do not declare the generative values of $\boldsymbol{\xi}_{z2}$ or $\boldsymbol{\xi}_{z3}$ in each of the matrices $\boldsymbol{\Xi}_z$ ($z \in [g]$) since they do not effect the respective densities due to each density having $\beta_z = 0$. Visualizations of S-I and S-II are provided Figures 1 and 2, respectively.

In order to evaluate the accuracy of the EEs, we repeat each of the studies S-I and S-II $R = 100$ times. From each repetition, we obtain the EE $\hat{\boldsymbol{\theta}}_n^{\langle l \rangle}$ for $l \in [R]$, where $\hat{\boldsymbol{\theta}}_n^{\langle l \rangle}$ contains $\hat{\pi}_{zn}^{\langle l \rangle}$, $\hat{\beta}_{zn}^{\langle l \rangle}$, $\hat{\kappa}_{zn}^{\langle l \rangle}$, and $\hat{\boldsymbol{\Xi}}_{zn}^{\langle l \rangle}$ for each $z \in [g]$. Using the sample of $R = 100$ EEs, we then compute the mean-squared errors (MSEs) of over some subsets of the parameter vector to their generative counterparts. That is, we compute the MSEs

$$\text{MSE}_\pi = (gR)^{-1} \sum_{z=1}^{g} \sum_{l=1}^{R} \left( \hat{\pi}_{zn}^{\langle l \rangle} - \pi_z \right)^2,$$

$$\text{MSE}_\kappa = (gR)^{-1} \sum_{z=1}^{g} \sum_{l=1}^{R} \left( \hat{\kappa}_{zn}^{\langle l \rangle} - \kappa_z \right)^2,$$

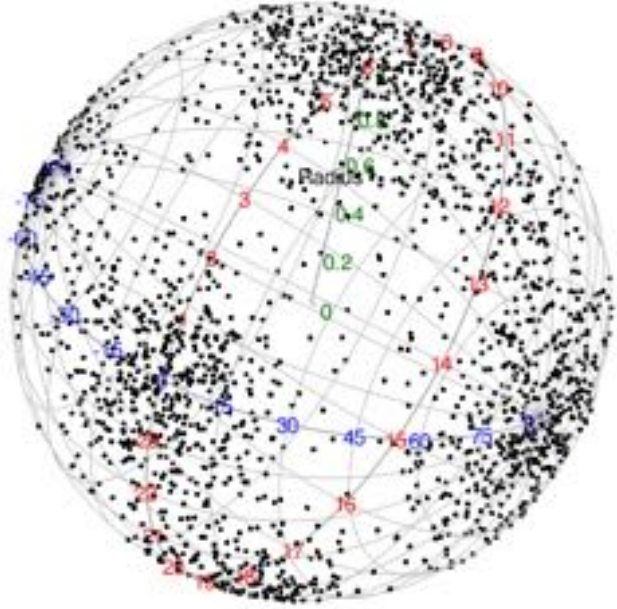

Figure 2: A single instance of an $n = 1000$ observations sample that is simulated under the $g = 6$ component mixture scenario of S-II.

and
$$\text{MSE}_{\boldsymbol{\xi}_{z1}} = R^{-1} \sum_{l=1}^{R} \left\| \hat{\boldsymbol{\xi}}_{z1n}^{\langle l \rangle} - \boldsymbol{\xi}_{z1} \right\|^2,$$

for each $z \in [g]$, for each of the studies S-I and S-II.

From S-I, we obtain the MSE results: $\text{MSE}_\pi = 0.000252$, $\text{MSE}_\kappa = 0.482$, $\text{MSE}_{\boldsymbol{\xi}_{11}} = 0.000790$, $\text{MSE}_{\boldsymbol{\xi}_{21}} = 0.000811$, and $\text{MSE}_{\boldsymbol{\xi}_{31}} = 0.000913$. From S-II, we obtain the MSE results: $\text{MSE}_\pi = 0.000131$, $\text{MSE}_\kappa = 2.80$, $\text{MSE}_{\boldsymbol{\xi}_{11}} = 0.000596$, $\text{MSE}_{\boldsymbol{\xi}_{21}} = 0.000550$, $\text{MSE}_{\boldsymbol{\xi}_{31}} = 0.000630$, $\text{MSE}_{\boldsymbol{\xi}_{41}} = 0.000532$, $\text{MSE}_{\boldsymbol{\xi}_{51}} = 0.000549$, and $\text{MSE}_{\boldsymbol{\xi}_{61}} = 0.000564$. All of the MSE values are relatively small compared to the absolute value of the parameter components being estimated. Thus, we can conclude that the EE based on the maximization of (5) is a sufficiently accurate method for parameter estimation in the simulation scenarios that were assessed.

## 4.2 Bayesian Information Criterion

We now assess the BIC-like criterion for choosing the number of mixture components $g$, which is presented in Section 3. Proposition 3 states that the BIC rule (17) asymptotically selects the most parsimonious model with respect to the maximization of the approximate log-likelihood function (5). Unfortunately, the result provides no finite sample guarantees. Furthermore, assumption (B2) of Proposition 3 is difficult to validate and thus we cannot be sure of the performance of rule (17) when applied to data. The following simulation study, S-III, is designed to assess the ability of rule (17) to correctly select the number of components $g$ in finite-data scenarios.

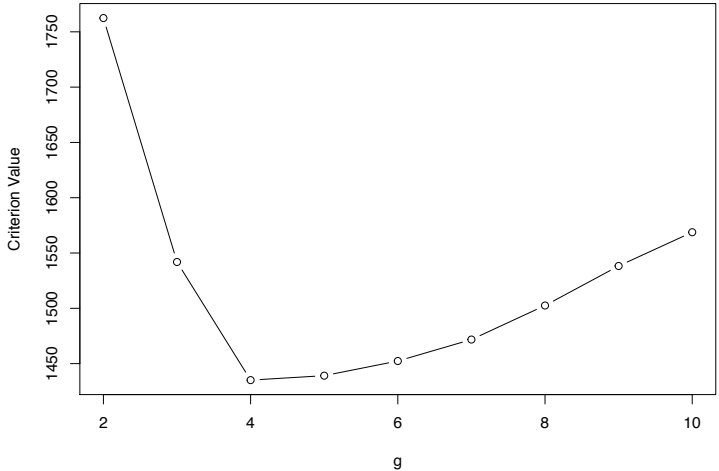

Figure 3: Average BIC-like criterion values of form (18) over $R = 100$ replications of S-III.

In S-III, we generate $n = 1000$ observations from a $g = 5$ component mixture of von Mises-Fisher distributions, with $\kappa_z = 10$ and $\pi_z = 1/3$ for all $z \in [g]$. The matrices $\mathbf{\Xi}_z$ are generated uniformly over the Stiefel manifold $V_3\left(\mathbb{R}^3\right)$, for each $z$, using the method described in Hoff (2009). For each $g \in \{2, \ldots, 10\}$, a $g$ component mixture of Kent distributions is estimated using the BSLM algorithm and the EE $\hat{\boldsymbol{\theta}}_n^{[g]}$ is noted. Rule (17) is then used to compute the optimal number of mixture components $\hat{g}_n$. In order to evaluate the accuracy of rule (17), we repeat S-III $R = 100$ times and note the number of times $\hat{g}_n = g$ for each $g \in \{2, \ldots, 10\}$.

From the $R = 100$ repetitions, we found that mixtures of Kent distributions with $g \in \{2, 3, 4, 5, 6, 7\}$ components were selected 9, 24, 31, 18, 9, and 9 times, respectively. A plot of the average BIC-like criterion value

$$-\tilde{L}\left(\hat{\boldsymbol{\theta}}_n^{[g]}; \mathbf{X}_n\right) + \mathrm{pen}_n\left(g\right), \tag{18}$$

for each $g \in \{2, \ldots, 10\}$, is displayed in Figure 3.

From the S-III results, we can observe that rule (17) appears to over penalize model complexity in the assessed scenario. The fact that we do not observe consistent selection of the true generative number of components $g = 5$ is not surprising, as the conclusion of Proposition 3 only applies asymptotically, and provides no guarantees in finite sample cases. Thus, more careful calibration of the penalization term $\mathrm{pen}_n\left(g\right)$ in (18) may be required for better finite-sample model selection performance.

### 4.3 Model-Based Clustering via Kent Distribution Mixtures

Thus far we have considered only the estimation of the parameter vector $\boldsymbol{\theta}$ from data $\mathbf{X}_n$ from a density estimation perspective. That is, we have considered only the case where we determine $\hat{\boldsymbol{\theta}}_n$ for the purpose of obtaining the $g$ component mixture of Kent distributions of form (3), which best fits the data, where the best-fitting distribution is defined as the one that minimizes the approximate log-likelihood function (5). We now consider an alternative interpretation of model (3) for the purpose of clustering heterogeneous data.

When conducting model-based clustering, we assume that our data $\mathbf{X}_n$ arises from a heterogeneous data generating process, made up of $g$ subpopulations, that can be best characterized by a mixture of Kent distributions, defined via the hierarchical construction that precedes Equation (3). That is, we assume that each datum $\boldsymbol{X}_i$ has a latent label $Z_i \in [g]$ that determines which of the $g$ different component densities $f_K(\boldsymbol{x}; \boldsymbol{\psi}_z)$ that it was generated from, for $z \in [g]$.

Let $\boldsymbol{x}$ be an arbitrary observation that is generated via a heterogeneous process, made up of that is best characterized by a $g$ component Kent distribution mixture with some known parameter vector $\boldsymbol{\theta}$. Suppose that we wish to estimate $z$ (i.e. the component from which $\boldsymbol{x}$ was generated). We can utilize the maximum a posteriori (MAP) rule

$$\hat{z} = \arg\max_{z \in [g]} \pi_z f_K(\boldsymbol{x}; \boldsymbol{\psi}_z), \tag{19}$$

which has optimal properties when the true data generating process matches the mixture model that is used (cf. Wasserman, 2004, Thm. 22.6).

Often, we do not know the parameter vector $\boldsymbol{\theta}$ that best characterizes the heterogeneity of the data $\mathbf{X}_n$. Furthermore, the normalizing constants (2) can cause computational difficulties when applying rule (19). As such, we can estimate can estimate $\boldsymbol{\theta}$ by $\hat{\boldsymbol{\theta}}_n$ and replace $f_K(\boldsymbol{x}; \boldsymbol{\psi}_z)$ with $\tilde{f}_K(\boldsymbol{x}; \boldsymbol{\psi}_z)$ in order to obtain the approximate plugin MAP rule

$$\hat{z} = \arg\max_{z \in [g]} \hat{\pi}_{nz} \tilde{f}_K\left(\boldsymbol{x}; \hat{\boldsymbol{\psi}}_{nz}\right). \tag{20}$$

If the number of subpopulations is unknown we can further estimate $g$ by $\hat{g}_n$, using rule (17). The effectiveness of rule (20) is assessed using simulation study S-IV, below.

In S-IV, we simulate the data as in S-I and also keep note of the latent label variable $z_i \in \{1, 2, 3\}$, for each $i \in [n]$. A $g = 3$ component mixture of Kent distribution is then estimated via our BSLM algorithm in order to obtain the EE $\hat{\boldsymbol{\theta}}_n$. Using rule (20), we then estimate each latent label by $\hat{z}_i$ for each $i \in [n]$. The adjusted Rand index (ARI) of Hubert and Arabie (1985) is then computed in order to assess the performance of rule (20). The process of simulation under study protocol S-IV and computation of the ARI are repeated $R = 100$ times and averaged in order to obtain an accurate measure of performance.

From the $R = 100$ repetitions, we found that the use of rule (20) in S-IV achieved an average ARI value of 0.939. In order to benchmark this value, we also performed clustering of the data that were generated under S-IV using the mixture of von Mises-Fisher distributions-based clustering algorithm of Hornik and Grun (2014). The clustering by mixture of von Mises-Fisher distributions achieved an average ARI value of 0.940. Thus, the two algorithms performed almost identically. We note that both of the average ARI values are on the high end of the scale (the ARI is bounded from above by one), which indicates that the clustering problem that was posed by S-IV was in fact an easy one. In the future, it would be interesting to assess the performance of rule (20) for clustering in more difficult scenarios.

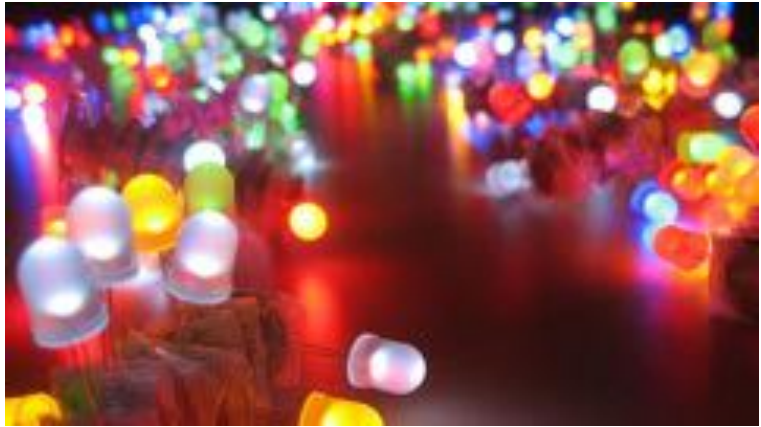

Figure 4: An example natural image that can be represented using the RGB color model.

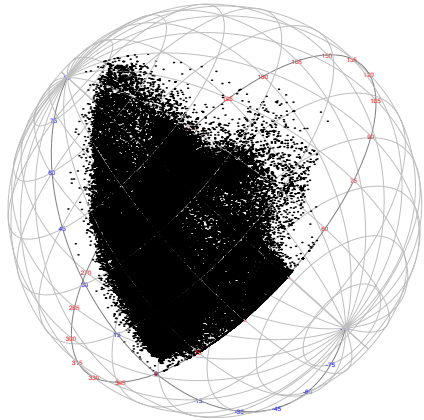

Figure 5: The RGB data from Figure 4, mapped onto the unit sphere $\mathbb{S}^2$.

## 5 Applications

### 5.1 Image Segmentation

The RGB (red-green-blue) color model is an additive model for representing the array of possible colors of light. Suppose that we have a natural image with $n$ pixels. When a natural image is stored under the RGB model, each of its pixels are represented as a vector in $\boldsymbol{y}_i^\top \in (\text{red}_i, \text{green}_i, \text{blue}_i)$, for $i \in [n]$, where each of the components of $\boldsymbol{y}_i$ represent the intensity of the named color, relative to the other colors in the vector.

Given data $\mathbf{y}_n = \{\boldsymbol{y}_i\}_{i=1}^n$, we can map each vector $\boldsymbol{y}_i$ onto the unit sphere $\mathbb{S}^2$ by taking the transformation $\boldsymbol{x}_i = \boldsymbol{y}_i / \|\boldsymbol{y}_i\|$, in order to obtain the sample $\mathbf{x}_n$. As an example, consider the image from Figure 4. Upon mapping each of its RGB pixels onto the unit sphere, we can visualize its $n = 129600$ data points in Figure 5.

Using the BIC-like criterion, a mixture of Kent distributions model with $g = 7$ components is selected in order to cluster the data from Figure 4. The result of the clustering is presented in Figure 6. We observe that the clustering ably separates out the background layer of the image and can distinguish between major color groups. This appears to be a successful test of the mixture of Kent distributions model as a segmentation tool and may lead to further development in this area of research.

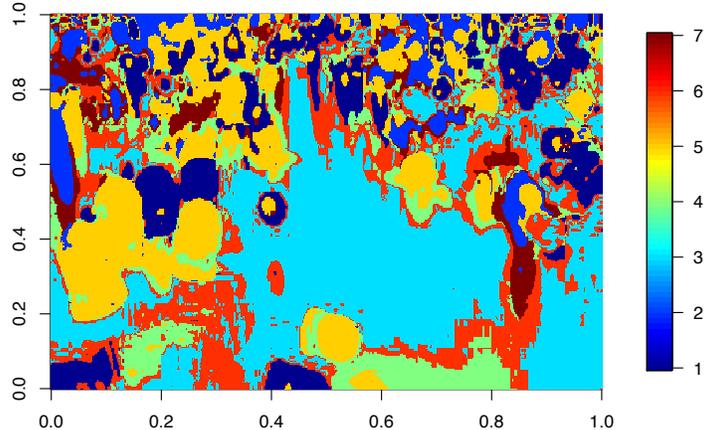

Figure 6: Clustering of data from Figure 4 using the mappings from Figure 5 and a $g = 7$ component mixture of Kent distributions model.

## 5.2 Neuroimaging

In neuroimaging studies involving MRIs (magnetic resonance images), it is common for practitioners to align their sample of scans to templates or "atlases" before further analyzing these samples. The alignment to these atlases allow the practitioners to target the study of various tissue types or neurological regions in a consistent manner across all of the scans in their samples simultaneously.

Many available atlases included a probability map that provides an indication regarding the likely tissue type at each voxel of the MRI scan. These maps often only provide three tissue types: cerebrospinal fluid (CSF), grey matter, and white matter. Thus, if we suppose that an MRI atlas probability map contains $n$ voxels, then each of the voxels can be represented by a probability vector $\bm{y}_i^\top = (\text{CSF}_i, \text{grey}_i, \text{white}_i)$, for $i \in [n]$, where each vector component indicates the probability that the voxel contains tissue of the labelled type.

Given all of the voxel probability maps from an atlas $\mathbf{y}_n = \{\bm{y}_i\}_{i=1}^n$, we can map each $\bm{y}_i$ onto the unit sphere $\mathbb{S}^2$ using the transformation $\bm{x}_i = \bm{y}_i / \|\bm{y}_i\|$, in order to obtain the sample $\mathbf{x}_n$. We demonstrate this process using a single slice from the ICBM 2009a Nonlinear Symmetric atlas of Collins *et al.* (1999), Fonov *et al.* (2009), and Fonov *et al.* (2011). In Figure 7, we visualize the T1-weighted image from the atlas set, which contains $n = 19219$ voxels. We then map the probability maps from the visualized slice in order to obtain the sample $\mathbf{x}_n$, which is plotted in Figure 8.

Using the BIC-like criterion, a mixture of Kent distributions model with $g = 8$ components is selected in order to cluster the data from Figure 8. The result of the clustering is presented in Figure 9. The clustering appears able to identify new classes of voxels that do not fit into the CSF, grey matter, and white matter split. These new classes of tissue types may lead to additional capacity for more specific and nuanced inference in MRI studies.

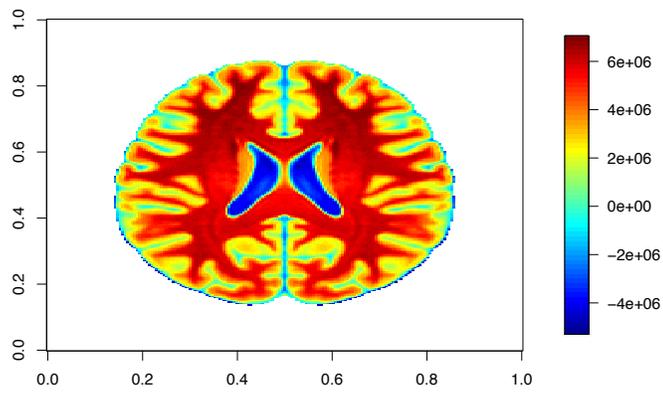

Figure 7: The $z = 95$ slice of the T1-weighted MRI from the ICBM 2009a Nonlinear Symmetric atlas set.

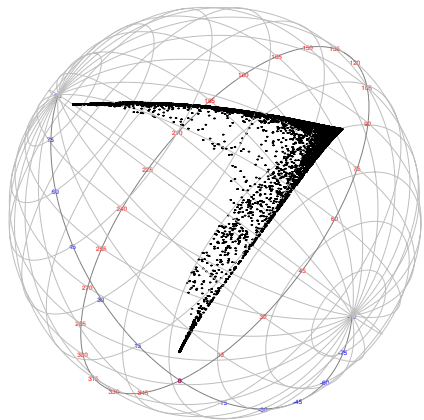

Figure 8: The probability map data from the voxels that are displayed in Figure 7, mapped onto the unit sphere $\mathbb{S}^2$.

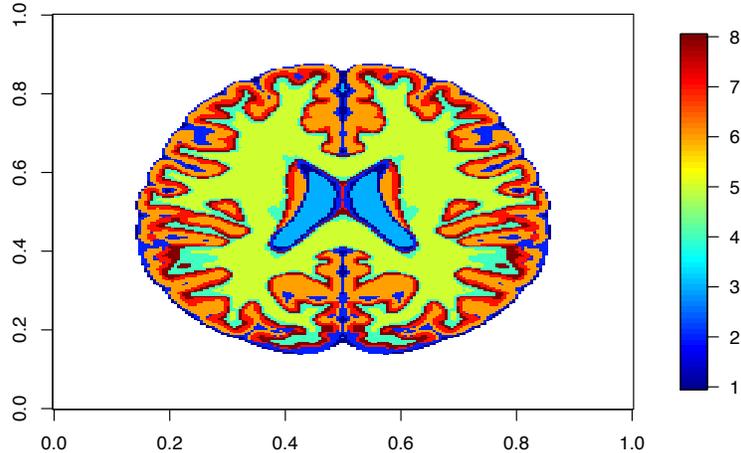

Figure 9: Clustering of data from Figure 8 using a $g = 8$ component mixture of Kent distributions model.

## 6  Conclusions

In this paper, we have presented a new algorithm for model-based clustering of data on the unit sphere $\mathbb{S}^2$. The approach is based on mixture of Kent distribution models. An asymptotic approximation to the Kent distribution probability density function is utilized in order to construct an approximate log-likelihood function, which is then used to conduct approximate ML estimation. The approximate ML estimator can be viewed as an EE in the sense of Amemiya (1985, Ch. 4). A BSLM algorithm is constructed in order to compute the approximate ML estimator from data. This algorithm utilizes some new algorithms for optimization on the Stiefel manifold, from Martin *et al.* (2016).

The BSLM algorithm monotonically increases the approximate log-likelihood function at each iteration. We proved that the BSLM algorithm is convergent under some mild regularity conditions. Furthermore, the approximate ML estimator is proved to be consistent. A BIC-like model selection rule is provided for the selection between different numbers of mixture components.

Via simulation studies, we demonstrated that the EE is accurate, and the BIC-like rule performed capably for the problem of model selection. Furthermore, we provided a model-based clustering rule that is based on the mixture of Kent distribution model and demonstrated that it compared favorably with a competitive method from Hornik and Grun (2014).

A pair of real data examples are used to demonstrate how the methodology could be utilized in practice. The first of these examples regards the segmentation of natural images, and the second of these examples pertains to the refinement of tissue segmentation of brain MRIs.

## Acknowledgements

The author is funded by Australian Research Council grant DE170101134.

# References

Absil, P.A., Baker, C.G., and Gallivan, K.A., 2007. Trust-region methods on Riemannian manifolds, *Foundations of computational mathematics*, 7, 303–330.

Absil, P.A., Mahony, R., and Sepulchre, R., 2008. *Optimization Algorithms on Matrix Manifolds*, Princeton: Princeton University Press.

Akaike, H., 1974. A new look at the statistical model identification, *IEEE Transactions on Automatic Control*, 19, 716–723.

Amemiya, T., 1985. *Advanced Econometrics*, Cambridge: Harvard University Press.

Banerjee, A., Dhillon, I.S., Ghosh, J., and Sra, S., 2005. Clustering on the unit hypersphere using the von Mises-Fisher distributions, *Journal of Machine Learning Research*, 6, 1345–1382.

Baudry, J.P., 2015. Estimation and model selection for model-based clustering with the conditional classification likelihood, *Electronic Journal of Statistics*, 9, 1041–1077.

Bingham, C., 1974. An andtipodally symmetric distribution on the sphere, *Annals of Statistics*, 2, 1201–1225.

Chen, X. and Yin, X., 2017. *NlcOptim: Solve Nonlinear Optimization with Nonlinear Constraints*.

Collins, D.L., Zijdenbos, A.P., Baare, W.F.C., and Evans, A.C., 1999. ANIMAL+INSECT: improved cortical structure segmentation, *in: IPMI Lecture Notes in Computer Science*, Springer, vol. 1613, 210–223.

de Leeuw, J., 1994. *Information Systems and Data Analysis*, Berlin: Springer, chap. Block-relaxation algorithms in statistics, 308–324.

Dempster, A.P., Laird, N.M., and Rubin, D.B., 1977. Maximum likelihood from incomplete data via the EM algorithm, *Journal of the Royal Statistical Society Series B*, 39, 1–38.

Edelman, A., Arias, T.A., and Smith, S.T., 1998. The geometry of algorithms with orthogonality constraints, *SIAM Journal on Matrix Analysis and Applications*, 20, 303–353.

Fisher, R.A., 1953. Dispersion on a sphere, *Proceedings of the Royal Society of London A*, 217, 295–305.

Fonov, V., Evans, A.C., Botteron, K., Almli, C.R., McKinstry, R.C., Collins, D.L., and the Brain Development Cooperative Group, 2011. Unbiased average age-approriate altases for pediatric studies, *NeuroImage*, 54, 313–327.

Fonov, V.S., Evans, A.C., McKinstry, R.C., Almli, C.R., and Collins, D.L., 2009. Unbiased nonlinear average age-approriate brain templates from birth to adulthood, *NeuroImage*, 47, S102.

Franke, J., Redenbach, C., and Zhang, N., 2016. On a mixture model for directional data on the sphere, *Scandinavian Journal of Statistics*, 43, 139–155.

Hoff, P.D., 2009. Simulation of the matrix Bingham-von Mises-Fisher distribution, with applications to multivariate and relational data, *Journal of Computational and Graphical Statistics*, 18, 438–456.

Hornik, K. and Grun, B., 2014. movMF: an R packaged for fitting mixtures of von Mises-Fisher distributions, *Journal of Statistical Software*, 58, 1–31.

Huang, W., Absil, P.A., Galivan, K.A., and Hand, P., 2016. ROPTLIB: an object-oriented C++ library for optimization on Riemannian manifolds, Tech. Rep. FSU16-14, Florida State University.

Huang, W., Gallivan, K.A., and Absil, P.A., 2015. A Broyden class of quasi-Newton methods for Riemannian optimization, *SIAM Journal of Optimization*, 25, 1660–1685.

Hubert, L. and Arabie, P., 1985. Comparing partitions, *Journal of Classification*, 2, 193–218.

James, I.M., 1976. *The Topology of Stiefel Manifolds*, Cambridge: Cambridge University Press.

Jennrich, R.I., 1969. Asymptotic properties of non-linear least squares estimators, *Annals of Mathematical Statistics*, 40, 633–643.

Kent, J.T., 1982. The Fisher-Bingham distribution on the sphere, *Journal of the Royal Statistical Society B*, 44, 71–80.

Keribin, C., 2000. Consistent estimation of the order of mixture models, *Sankhya A*, 62, 49–65.

Leroux, B.G., 1992. Consistent estimation of a mixing distribution, *Annals of Statistics*, 20, 1350–1360.

Mardia, K.V., 1975. Statistics of directional data, *Journal of the Royal Statistical Society B*, 37, 349–393.

Mardia, K.V. and Jupp, P.E., 2000. *Directional Statistics*, Chichester: Wiley.

Martin, S., Raim, A.M., Huang, W., and Adragni, K.P., 2016. ManifoldOptim: An R interface to the ROPTLIB library for manifold optimization, *arXiv:1612.03930*.

McLachlan, G.J., 1988. On the choice of starting values for the EM algorithm in fitting mixture models, *The Statistician*, 37, 417–425.

McLachlan, G.J. and Krishnan, T., 2008. *The EM Algorithm And Extensions*, New York: Wiley, 2nd ed.

McLachlan, G.J. and Peel, D., 2000. *Finite Mixture Models*, New York: Wiley.

Meyer, R.R., 1976. Sufficient conditions for the convergence of monotonic mathematical programming algorithms, *Journal of computer and system sciences*, 12, 108–121.

Nocedal, J. and Wright, S.J., 2006. *Numerical Optimization*, New York: Springer.

Olver, F.W.J., Lozier, D.W., Boisvert, R.F., and Clark, C.W., eds., 2010. *NIST Handbook of Mathematical Functions*, Cambridge: Cambridge University Press.

Peel, D., Whiten, W.J., and McLachlan, G.J., 2001. Fitting mixtures of Kent distributions to aid in joint set identification, *Journal of the American Statistical Association*, 96, 56–63.

R Core Team, 2016. *R: a language and environment for statistical computing*, R Foundation for Statistical Computing.

Razaviyayn, M., Hong, M., and Luo, Z.Q., 2013. A unified convergence analysis of block successive minimization methods for nonsmooth optimization, *SIAM Journal of Optimization*, 23, 1126–1153.

Schwarz, G., 1978. Estimating the dimensions of a model, *Annals of Statistics*, 6, 461–464.

Sra, S. and Karp, D., 2013. The multivariate watson distribution: maximum-likelihood estimation and other aspects, *Journal of Multivariate Analysis*, 114, 256–269.

van der Vaart, A., 1998. *Asymptotic Statistics*, Cambridge: Cambridge University Press.

Wasserman, L., 2004. *All Of Statistics: A Concise Course In Statistical Inference*, New York: Springer.

Yamaji, A. and Sato, K., 2011. Clustering of fracture orientations using a mixed Bingham distribution and its application to paleostress analysis from dike or vein orientations, *Journal of Structural Geology*, 33, 1148–1157.

Yang, M.S., Chang-Chien, S.J., and Hung, W.L., 2016. An unsuperised clustering algorithm for data on the unit hypersphere, *Applied Soft Computing*.

Zhou, H. and Lange, K., 2010. MM algorithms for some discrete multivariate distributions, *Journal of Computational and Graphical Statistics*, 19, 645–665.